\documentstyle[preprint,aps]{revtex}  % for preprint form  (RevTex 3.0)
\newcommand{\ul}[1]{\underline{#1}}
\newcommand{\E}{{\cal E}}

\tighten       %Gives single-space (RevTex 3.0)

\begin{document}
%prints PACS numbers in \pacs{ }
\draft

\title{Impact of interface roughness on perpendicular transport and domain
formation  in superlattices}
\author{Andreas Wacker and Antti-Pekka Jauho}
\address{Mikroelektronik Centret,
Danmarks Tekniske Universitet, DK-2800 Lyngby, Denmark}

\date{Contribution to ICSMM-9 in Li{\`e}ge, Belgium, July 1996\\
to appear in Superlattices and Microstructures}
\maketitle

%**********************************************************************%
%
\begin{abstract}
A microscopic calculation of the perpendicular current in 
doped multiple quantum wells is presented. 
Interface roughness is shown  to affect the resonant transitions 
as well as to cause a nonresonant
background current. The theoretical characteristics exhibit
several branches due to the formation of electric field domains 
in quantitative agreement with experimental data.
\end{abstract}

\section{Introduction}
The electric transport in semiconductor superlattices is usually dominated
by resonances between the localized energy levels inside the wells
resulting in peaks  in the current-field relation.
This may yield complicated current-voltage 
characteristics exhibiting may branches  due to the formation of 
domains of different electric field
inside the sample (see, e.g., \cite{KWO95} and references therein).
These experimental features could be qualitatively reproduced  
by theoretical models
combining rate equations for the transport between the wells
and Poisson's equation \cite{PRE94,BON94}.
While these approaches assume interwell transition rates which are 
either fitted
or obtained from phenomenological models, we have recently
proposed a way to calculate the transport microscopically \cite{WACpa}.
We obtained good quantitative agreement with the experimental
data of Ref.~\cite{HEL91} for highly doped samples, where
the scattering from ionized impurities causes a strong broadening 
of the levels.
Here we consider the lower doped sample used in \cite{KWO95,SCH96a}
consisting of a 40 period GaAs/AlAs superlattice (barrier width
$b=4$ nm, well width $w=9$ nm, period $d=b+w$, 
doping $N_D=1.5\times 10^{11}/$cm$^2$ per well, 
cross section $A=1.13\times 10^{-4}$ cm$^2$) and investigate the 
impact of interface roughness which both contributes to the broadening
and causes nonresonant transitions between the wells.  

\section{The basic transport model}
In the case of  weakly coupled  quantum wells the appropriate basis set 
is a product of Wannier functions $\Psi^{\nu}(z-nd)$ of subband $\nu$
localized in well $n$, and plane waves $e^{i\ul{k}\cdot \ul{r}}$.
Here the $z$ direction is the growth direction and
$\ul{k},\ul{r}$ are vectors within the $(x,y)$ plane.
Restricting  ourselves to the lowest two minibands (denoted by $a$ and $b$) 
and coupling between neighbouring wells the Hamiltonian in the presence
of an electric field $F$  is given by
$\hat{H}=\hat{H}_0+\hat{H}_1$, where
\begin{eqnarray*}
\hat{H}_0&=&\sum_{n,\ul{k}} \left[
E_n^{a}(\ul{k})a_n^{\dag}(\ul{k})a_n(\ul{k})
+E_n^{b}(\ul{k})b_n^{\dag}(\ul{k})b_n(\ul{k})
-eF(R_0^{ab}a_n^{\dag}(\ul{k})b_n(\ul{k})
+R_0^{ba}b_n^{\dag}(\ul{k})a_n(\ul{k}))
\right]\label{Eqham1}\\
\hat{H}_1&=&\sum_{n,\ul{k}} \left[
T_1^a a_{n+1}^{\dag}(\ul{k})a_n(\ul{k})
+T_1^b b_{n+1}^+(\ul{k})b_n(\ul{k})
-eFR^{ab}_1 a_{n+1}^{\dag}(\ul{k})b_n(\ul{k})
-eFR^{ba}_1 b_{n+1}^{\dag}(\ul{k})a_n(\ul{k}) \right]
\end{eqnarray*}
with $E_n^{\nu}(\ul{k})=E^{\nu}+\hbar^2k^2/(2m_w)-eFdn$ 
($m_w$ is  the effective mass  in the well), the couplings 
$R_h^{\nu'\nu}=\int dz \Psi^{\nu'}(z-hd)z\Psi^{\nu}(z)$, and the
miniband width $4|T_1^{\nu}|$ of subband $\nu$. Diagonalizing the Hamiltonian
$\hat{H}_0$ leads  to renormalized
coefficients in $\hat{H}_0$ and $\hat{H}_1$\cite{KAZ72} 
which we use in the following.

We calculate the Wannier functions in a Kronig-Penney-type model.
Following Ref.~\cite{BRO90} we model the nonparabolicity
by an energy dependent effective mass $m(E)=m_c(1+(E-E_c)/E_g)$,
where $m_c$ is the effective mass at the conduction band minimum of 
energy $E_c$, and $E_g$ is the energy gap. Then the usual connection rules
hold for the envelope function provided that the momentum matrix element
$P$ between the conduction and valence band states is identical in 
both materials. We use the values\cite{BRO90} $m_c^{{\rm GaAs}}=0.067 m_e$,
$m_c^{{\rm AlAs}}=0.15 m_e$, $E_g^{{\rm GaAs}}=1.52$ eV, 
$E_g^{{\rm AlAs}}=3.13$ eV, and the conduction band 
discontinuity $\Delta E_c=1.06$ eV.
These parameters yield a relation $E(k)=E_c+\hbar^2k^2/(2m(E))$ 
which is in excellent agreement with the band structure of 
AlAs\cite{SCH85a} for the energies of interest.\footnote{Within this 
approach $P=\hbar\sqrt{E_g/(2m_c)}$ is slightly
different for the two materials in contrast to the assumption. Furthermore,
the envelope functions for different energies are not orthogonal as the 
effective Hamiltonian is energy dependent. However, the
overlap is small and we neglect these complications.}
We obtain the  coefficients 
$E^a=47.1$ meV, $E^b=$ 176.6 meV,
$T_1^a=-0.0201$ meV, $T_1^b=0.0776$ meV,
$R_1^{ba}=2.66\times 10^{-4}d$, and $R_0^{ba}=-0.149d$.

For small couplings between the wells and fast intersubband relaxation 
the current from subband 
$\nu$ in well $n$ to subband $\mu$ in well $n+1$
is given by the following expression\cite{MAH90}:
\begin{eqnarray}
I_{n\to n+1}^{\nu \to \mu}=2e\sum_{\ul{k}',\ul{k}}
|H_{(n+1)\ul{k},n\ul{k}'}^{\mu,\nu}|^2
\int_{-\infty}^{\infty} \frac{d\E}{2\pi \hbar}
A_n^{\nu}(\ul{k}',\E)
\cdot A_{n+1}^{\mu}(\ul{k},\E-\Delta\mu)
\left[n_F(\E)-n_F(\E-\Delta\mu)\right] \label{EqJ}\, .
\end{eqnarray}
Here $A_n^{\nu}(\ul{k},\E)$ is the spectral function 
of subband $\nu$ in well number $n$ and 
$n_F(\E)=(1+e^{\beta\E})^{-1}$ is the Fermi function.
The energy $\E$ is measured with respect to the 
electrochemical potential $\mu_n$ in well $n$
yielding an implicit dependence of $A_n^{\nu}(\ul{k},\E)$ on $\mu_n$.
We determine $\mu_n$ from the  local  electron density 
$n_n=\sum_{\ul{k}}\int_{-\infty}^{\infty} d\E
n_F(\E)A_n^{a}(\ul{k},\E)/(\pi A)$. Then the difference
$\Delta\mu=\mu_{n+1}-\mu_n$ is equal to $-eFd$ for $n_n=n_{n+1}$.
We obtain $A_n^{\nu}(\ul{k},\E)$  
in equilibrium from  the retarded self-energy 
$\Sigma_{n}^{\nu\,{\rm ret}}(\ul{k},\E)$ neglecting the coupling to the
other wells. 
In Ref.~\cite{WACpa} we have calculated 
the self-energy  for scattering from the screened potential of 
ionized impurities within the self-consistent
single-site-approximation. 
As an additional  contribution to the self-energy we 
study here the impact of interface roughness.

\section{Modelling of interface roughness}
We consider an interface located at $z=z_0$ exhibiting thickness 
fluctuations $\xi(\ul{r})$ of the order of
$\pm \eta$ (we use $\eta=2.8$ {\AA} which is one monolayer of GaAs). 
We assume  $\langle \xi(\ul{r}) \rangle_r=0$ and the correlation
\begin{equation}
\langle \xi(\ul{r})\xi(\ul{r}') \rangle_r
=\alpha \eta^2\exp(-|\ul{r}-\ul{r}'|/\lambda)\label{Eqexpkorr}\, .
\end{equation} 
This can be  motivated in the following way:
At a given point $\ul{r}$ there is an island of thickness $\eta$  with
a probability $\alpha$  (we use an average coverage  $\alpha=0.5$). 
Therefore 
$\langle \xi(\ul{r})\xi(\ul{r})\rangle_r=\alpha \eta^2$. Provided 
the island extends from $\ul{r}$ to $\ul{r}'$ we assume  a 
constant probability to find a further neighbouring atom 
beyond $\ul{r}'$ yielding the exponential in Eq.~(\ref{Eqexpkorr}).
Following Ref.~\cite{DHA90} we model the 
additional potential by a $\delta$-function at the perfect interface
$U(\ul{r},z)=\xi(\ul{r})\Delta E_c \delta(z-z_0)$ and obtain
\begin{eqnarray}
\hat{H}^{\rm rough}=\frac{1}{A}
&\sum_{\ul{k},\ul{p},h}&\left[
U^{aa}_{h}(\ul{p})a^{\dag}_{n+h}(\ul{k}+\ul{p})a_{n}(\ul{k})
+U^{bb}_{h}(\ul{p})b^{\dag}_{n+h}(\ul{k}+\ul{p})b_{n}(\ul{k})\right. 
\nonumber \\
&&\left.
+U^{ba}_{h}(\ul{p})b^{\dag}_{n+h}(\ul{k}+\ul{p})a_{n}(\ul{k})
+U^{ab}_{h}(\ul{p})a^{\dag}_{n+h}(\ul{k}+\ul{p})b_{n}(\ul{k})
\right]\label{Eqhamrough}
\end{eqnarray}
with the matrix elements $U^{\nu\mu}_h(\ul{p})=
\int d^2r e^{-i\ul{p}\cdot\ul{r}}\Delta E_c
\left[\xi(\ul{r})\Psi^{\nu\, *}(z_0-hd)
\Psi^{\mu}(z_0)\right]$.
The elements $U_1^{\nu\mu}$ contribute to the current from 
one well to the next
via Eq.~(\ref{EqJ}). For weakly coupled wells 
$U_2^{\nu\mu}, U_3^{\nu\mu}, \ldots$  are small and are neglected in the following.
The elements $U_0^{\nu\mu}$ result in scattering within the wells.
We calculate their contribution to the self-energy within the
self-consistent Born-Approximation
\begin{equation}
\Sigma_{{\rm rough}}^{a\, {\rm ret}}(\ul{k},\E)
=\frac{2}{A^2}\sum_{\ul{k}'}
\frac{|U^{aa}_0(\ul{k}-\ul{k}')|^2}
{\E+\E_n^F-E_{k'}-\Sigma_{n}^{a\, {\rm ret}}(\ul{k}',\E)}\, ,
\label{Eqsigmarough}
\end{equation}
where the factor 2 takes into account the two interfaces per well.
The calculation for the subband $b$ is performed in the same way.
With (\ref{Eqexpkorr}) we obtain the matrix element
\begin{equation}
|U^{aa}_0(\ul{p})|^2=A \Delta E_c^2
|\Psi_a(z_0)|^4
\frac{2\pi\alpha \eta^2 \lambda^2}{\left(1+(p\lambda)^2\right)^{3/2}}
\label{EqUqexp}
\end{equation}
In contrast to this the usual choice (see, e.g., \cite{DHA90}) 
$\langle \xi(\ul{r})\xi(\ul{r}') \rangle_r=
\alpha \eta^2\exp(-|\ul{r}-\ul{r}'|^2/\lambda^2)$ yields
$|U^{aa}_0(\ul{p})|^2\sim \exp(-\lambda^2p^2/4)$.
In order to compare both expressions we consider
the  matrix element for scattering at a single
circular island of thickness $\eta_i$ and radius $\lambda_i$ which is given by
$U_0^{aai}(\ul{p})=2\pi \eta_i \Delta E_c |\Psi_a(z_0)|^2\lambda_i
J_1(p\lambda_i)/p$,
where $J_1$ is the Bessel function of the first kind.
Now there are $\alpha A/(\pi \lambda^2)$ islands and we assume that
negative and positive $\eta_i$ cancel each other. Then 
$|U^{aa}(\ul{p})|^2=\sum_i |U_0^{aai}(\ul{p})|^2\sim p^{-3}$ for large $p$
as in Eq.~(\ref{EqUqexp}) with a prefactor differing
by $2/\pi$. This indicates that the choice (\ref{Eqexpkorr})
captures an essential part of physics not contained in the 
usual choice.

Furthermore, the single island result reflects 
the quality of the Born-approximation for the self-energy. The 
analysis of a $\delta$-potential reveals that diagrams containing
multiple scattering at a single island
become important if the product 
$U(p) m/(2\hbar^2)$ is larger than 1. 
Using the value $U_0^{aai}(\ul{0})$ this yields 
$2\hbar^2/(\pi m\lambda^2)<\eta\Delta E_c |\Psi_a(z_0)|^2$.
(For our parameters  $\lambda>15$ nm.)
This expression can be interpreted easily:
The right hand side is the energy the electron gains 
if it is located  at an island exhibiting a larger well width. 
The left hand side is the
quantization energy associated with a length $\lambda$.
The localization of the wavefunction  takes place if the energy gain 
due to the larger well width dominates the cost due to the localization.
In this case higher order diagrams become important.

\section{Results and discussion}
Summing the contributions between different subbands given by Eq.~(\ref{EqJ}) 
we obtain the current $I_{n\to n+1}(F,n_n,n_{n+1})$ depending on the  
field and the electron densities.
Fig.~1a shows $I_{n\to n+1}(F,N_D,N_D)$ exhibiting  
a first maximum $I=0.33$ mA at $eFd=10$ meV where the current is
dominated by  $T_1^a$ transitions and a second  peak $I=1.8$ mA at
$eFd=140$ meV due to the $a\to b$ resonance.
The  interface roughness has two implications:
Firstly the maxima are broadened and slightly shifted  
due to the contribution to the self-energies  via
Eq.~(\ref{Eqsigmarough}). Secondly there is a 
nonresonant current due to the $U_1^{aa}$ transitions 
dominating the behaviour between the maxima.

In order to study the domain formation we consider
effective fields $F_i$ between  wells $i$ and $i+1$ fulfilling 
the discretized Poisson equation 
$\epsilon_r\epsilon_0(F_i-F_{i-1})=e(n_i-N_D)$. Then  the
local currents  $I_{i\to i+1}(F_i,n_i,n_{i+1})$ have to be equal
for all $i$ in the steady state.
The total voltage across the superlattice is $U=d\sum_{i=1}^NF_i$.
Using the boundary conditions $n_1=2N_D$, $n_{N+1}=2N_D$ we obtain the 
current-voltage characteristic shown in Fig.~1b which exhibits 
the usual sequence of branches \cite{KWO95,PRE94}.
In contrast to previous theoretical results the maximal current of the 
branches (90 $\mu$A) is significantly lower than the height of the 
first resonance.
It is  almost independent of the boundary conditions, which  
mainly affect the voltage where the leftmost homogeneous branch
breaks up. The maximal current of  90 $\mu$A
is between the two experimental results\cite{SCH96a} 
60 $\mu$A and 130 $\mu$A for samples \#1 and \#2, respectively, which both
have the nominal sample parameters used here.
We can reproduce the  difference in current between both samples
by assuming barrier widths which differ by just one monolayer,
thus suggesting  a possible explanation for the experimental discrepancy.
The minimal currents of the branches (26 $\mu$A) seem to be  
lower than the experimental values. They depend strongly on the nonresonant 
currents, e.g., we obtain the value 10 $\mu$A ignoring transitions via $U_1$.
This indicates that we have probably underestimated the 
nonresonant current.
The slope of the branches is  steeper in our calculation
than found experimentally; this could be due to
some additional  contact resistance of the order of  1 k$\Omega$ which is  
not included in our calculation.

In conclusion, we have presented a microscopic model for sequential tunnelling
in weakly coupled quantum wells. The model  describes the
domain formation in superlattices quantitatively with no fitting parameters.
Nonresonant transitions due to interface roughness have been shown to 
strongly affect the current between the resonances and the extent 
of the current branches.
%\bibliography{ref}

\begin{thebibliography}{10}

\bibitem{KWO95}
S.~H. Kwok, H.~T. Grahn, M. Ramsteiner, K. Ploog, F. Prengel, A. Wacker, E.
  Sch{\"o}ll, S. Murugkar, and R. Merlin, Phys.~Rev.~B {\bf 51},  9943  (1995).
%S.~H. Kwok {\it et~al.}, Phys.~Rev.~B {\bf 51},  9943  (1995).

\bibitem{PRE94}
F. Prengel, A. Wacker, and E. Sch{\"o}ll, Phys.~Rev.~B {\bf 50},  1705  (1994),
  erratum in {\bf 52}, 11518 (1995).

\bibitem{BON94}
L.~L. Bonilla, J. Gal{\'a}n, J.~A. Cuesta, F.~C. Mart\'{\i}nez, and J.~M.
  Molera, Phys.~Rev.~B {\bf 50},  8644  (1994).
%L.~L. Bonilla {\it et~al.}, Phys.~Rev.~B {\bf 50},  8644  (1994).

\bibitem{WACpa}
A. Wacker and A.-P. Jauho, Physica~Scripta, in print.

\bibitem{HEL91}
P. Helgesen, T.~G. Finstad, and K. Johannessen, J.~Appl.~Phys. {\bf 69},  2689
  (1991).

\bibitem{SCH96a}
G. Schwarz, A. Wacker, F. Prengel, E. Sch{\"o}ll, J. Kastrup, H.~T. Grahn, and
  K. Ploog, Semicond.~Sci.~Technol. {\bf 11},  475  (1996).
%G. Schwarz {\it et~al.}, Semicond.~Sci.~Technol. {\bf 11},  475  (1996).

\bibitem{KAZ72}
R.~F. Kazarinov and R.~A. Suris, Sov.~Phys.~Semicond. {\bf 6},  120  (1972).

\bibitem{BRO90}
G. Brozak, E.~A. de~Andrada~e Silva, L.~J. Sham, F. DeRosa, P. Miceli, 
S.~A. Schwarz, J.~P. Harbison, L.~T. Florez, and J.~S. Allen, 
Phys.~Rev.~Lett. {\bf 64},  471  (1990).
%G. Brozak {\it et~al.}, Phys.~Rev.~Lett. {\bf 64},  471  (1990).

\bibitem{SCH85a}
J.~N. Schulman and Y.-C. Chang, Phys.~Rev.~B {\bf 31},  2056  (1985).

\bibitem{MAH90}
G.~D. Mahan, {\em Many-Particle Physics} (Plenum, New York, 1990).

\bibitem{DHA90}
I. Dharssi and P.~N. Butcher, J.~Phys.: Condens.~Matter {\bf 2},  4629  (1990).

\end{thebibliography}
%\bibliographystyle{aw_pr}

\begin{figure}
\vspace*{5.5cm}
\includegraphics{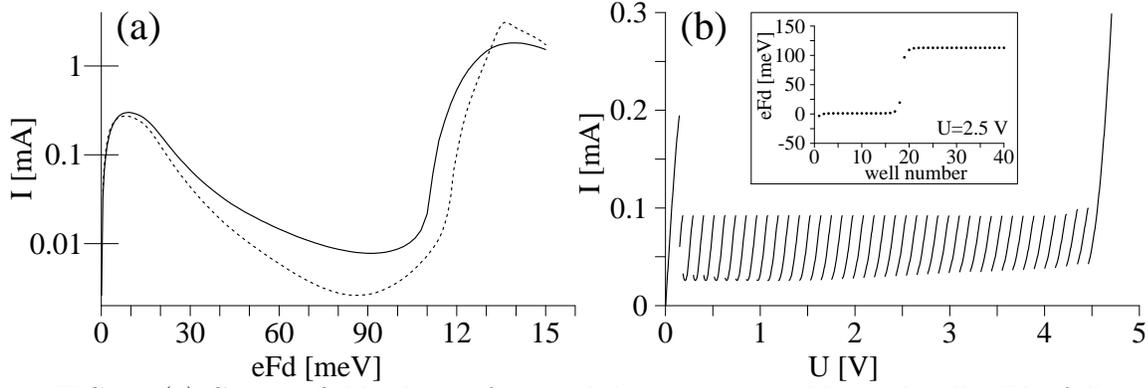}
\caption[fig1]{(a) Current-field relation for equal densities in 
neighboured wells. The full and dotted lines are calculated with and  without 
interface roughness, respectively (assuming $\lambda=7$ nm).
(b) Current-voltage characteristics for a superlattice consisting of 
$N=40$ wells for voltage sweep-up. The inset gives 
the field profile at $U=2.5$ V.}
\end{figure}

\end{document}